\documentclass[aps,pre,twocolumn,showpacs,amssymb]{revtex4-1}

\usepackage[dvips]{graphicx}
\usepackage{dcolumn}
\usepackage{bm}
\usepackage{color}
\def\be{\begin{equation}}
\def\en{\end{equation}}
\def\bea{\begin{eqnarray}}
\def\ena{\end{eqnarray}}
\def\bec{\begin{equation}\begin{array}{rcl}}

\def\p{\partial}
\def\ep{\epsilon}
\def\gs{\gtrsim}
\def\ls{\lesssim}

\newcommand{\av}[1]{\langle{#1}\rangle}
\newcommand{\AV}[1]{\bigg \langle{#1}\bigg \rangle}
\newcommand{\bi}[1]{\mbox{\boldmath$#1$}}

\def\hrij{\hat{\bi r}_{ij}}

\begin{document}
\title{
Formation of double   glass 
in binary mixtures of anisotropic particles:\\ 
Dynamic heterogeneities in rotations and displacements  
}  
\author{Kyohei Takae and Akira Onuki}
\affiliation{Department of Physics, Kyoto University, Kyoto 606-8502, Japan}


\date{\today}

\begin{abstract} 
We study glass behavior in a 
mixture of  elliptic  and circular particles 
in two dimensions at low temperatures 
using  an orientation-dependent    Lennard-Jones 
potential.   The ellipses  have a mild  aspect ratio ($\sim 1.2)$ and 
tend to align at low temperatures, 
while  the circular particles play the role of 
impurities disturbing the ellipse orientations  
at a concentration of $20\%$. 
These impurities  have a size smaller than that of the ellipses 
and attract them  in the homeotropic 
alignment. As a result,  the coordination number around each 
impurity is mostly  five or four 
in glassy states.    We  realize double glass,  
where both the orientations and  the positions are   
 disordered but still hold  mesoscopic order. 
We find a strong heterogeneity  in the 
 flip motions of the ellipses, which sensitively 
depends on the  impurity clustering. 
In our model, a small fraction of 
the ellipses still undergo flip  motions relatively 
rapidly  even at low temperatures.
 In contrast, the  non-flip rotations 
(with angle changes not close to $\pm \pi$) 
are mainly caused by  the cooperative  
configuration changes involving many  
particles. Then, there  arises a long-time  heterogeneity  
in   the   non-flip rotations  
closely correlated with the dynamic 
heterogeneity  in displacements. 
\end{abstract}

\maketitle

\section{Introduction} 

Much attention has been 
paid to various types of 
 glass transitions, where  the structural  relaxations become extremely  slow 
with lowering the temperature $T$ \cite{Ang,Binder}. 
In experiments, colloidal particles  can be spherical, but    
 real molecules are mostly nonspherical.  The  
translational and rotational  diffusion constants 
 have  thus   been measured 
in molecular systems 
near the  glass transition \cite{Silescu}. 
Using  generalized  mode-coupling theories,       
some authors    \cite{Rolf,Sch,Gotze} have studied 
the coupled translation-rotation dynamics 
to predict  translational glass and  orientational glass.  
Theoretically, for double glass \cite{Sch},  
 the translational and orientational  degrees of freedom 
   can be  simultaneously   arrested   
at the same temperature.  
In real systems, the  molecular rotations 
sensitively depend on many parameters 
including the molecular shapes,  the density  
 and the concentration (for mixtures).

Mixtures of    anisotropic particles 
with   mild differences in  sizes and shapes 
such as (KCN)$_x$(KBr)$_{1-x}$   
 form a cubic crystal without 
orientational order (plastic solid) at relatively 
high $T$.  With further lowering  $T$, 
they  undergo a structural phase transition 
in  dilute cases  and become  orientational  
 glass   in nondilute cases   \cite{ori}, 
where the crystal structure is preserved. 
On the other hand, if the two species have  
significantly different   sizes or shapes, 
translational  glass without crystal order 
can emerge from liquid   at low $T$. For 
 rodlike molecules with 
relatively large aspect ratios, 
liquid crystal phase transitions 
occur with lowing $T$, but their glass transitions have not yet 
well understood. 
In  recent  experiments on    colloidal 
ellipsoids   in monolayers, the    aspect ratio  
was 6 \cite{China} or 2.1 \cite{Mishra}  
with considerable size dispersities. In glassy states,    
these ellipsoids exhibited  
mesoscopic nematic or smectic order. 

Molecular dynamics simulations 
have also  been performed  on glass-forming fluids 
 composed of  anisotropic particles. 
They can be one-component fluids 
with  a complex internal structure. 
Examples are  methanol \cite{Klein}, 
ortho-terphenyl methanol(OTP)\cite{Lewis,Debe}, 
and  fluids of asymmetric dumbbells     
   \cite{Kob1,Michele}.    
There are various kinds  of 
two-component glass-formers. 
The simplest example is   a mixture of 
two species  of symmetric 
dumbbells \cite{Chong1,Moreno,Chong}. 
Recently, we  studied a mixture of 
spheroidal  and spherical particles 
to examine the orientational glass 
using an orientation-dependent potential \cite{EPL}.

The physical picture of double glass 
is thus very complex. To give a clear example, 
 we consider a mixture of 
elliptic particles with  a mild aspect ratio 
  and smaller circular  particles (impurities).
We assume  orientation-dependent repulsive and attractive 
interactions, where    the attractive part 
is  between the ellipses and the impurities. 
Then,  the  impurities can   strongly disturb  the  
orientations and the positions of  the surrounding ellipses. 
This is analogous to hydration of ions by 
surrounding water molecules \cite{Is,water}. 
If the impurity concentration 
$c$  is increased from zero, orientational 
domains and crystalline grains 
of the ellipses are gradually fragmented 
and disordered \cite{EPL}.   In this paper, 
we   realize double glass at low $T$ 
 at an impurity  concentration of $20\%$.

To  produce glassy states, we slowly quench the mixture 
from liquid. In this situation,  
we encounter  impurity clustering 
or aggregation at low $T$, 
which often results in small crystalline domains 
of impurities \cite{An,Lek}.
In our model, this tendency is  considerably 
suppressed by  the above-mentioned impurity-ellipse 
attractive interaction.   
Nevertheless, the impurity distribution is 
still mesoscopically heterogeneous, leading to 
 a mesoscopic heterogeneity 
in  the rotational motions.  
We shall see that   some  fraction 
of the ellipses still  rotate 
under  weak   constraints even at  low $T$. 
Furthermore, if anisotropic particles have 
the elliptic symmetry (the spheroidal one in three dimensions), 
they   can undergo flip (turnover) motions with $\pm \pi$ angle 
changes  \cite{Kob1,Klein,Michele}. 
 These flip motions can occur thermally 
for  mild aspect ratios, while they are   sterically  
 hindered by  the surrounding particles   
for large aspect ratios. 
 Thus, we expect   a wide range of the rotational activity 
 for mild aspect ratios.

We shall find 
 marked orientational and positional  heterogeneities 
on mesoscopic scales in glass.  
Such heterogeneous patterns have been 
visualized in various model systems 
\cite{Shintani,Hamanaka,Tanaka}.  
First, there arises  a mesoscopic  heterogeneity of the flip 
motions correlated with  the impurity clustering. 
Second, the positional configuration changes 
cause  non-flip rotations of the ellipses, which   are the origin 
of the long-time decay of the rotational correlation functions 
$G_\ell(t)$ of  even $\ell$ 
\cite{Klein,Michele,Kob1,Chong,Debe,Chong1,Moreno}. 
It follows   a  dynamic heterogeneity 
of the long-time non-flip rotations  correlated 
with the  dynamic heterogeneity in displacements 
or bond breakage \cite{Donati,Glotzer,YO,Shiba}.

The organization of this paper is as follows. 
In Sec.II,  we will explain our simulation  model 
and  method.  
In Sec.IIIA, we will present simulation results 
on the  heterogeneities in the orientations 
and the positions.  
 In Sec.IIIB, the time-correlation functions 
will be examined.
In Sec.IIIC, the angular and translational mean-square 
displacements will be calculated. 
In Sec.IIID, we will introduce the flip number 
for each ellipse in a  time interval 
and study its heterogeneity. 
In Sec.IIIE, time-development of a configurational change 
with  large displacements and/or  large 
angle changes will be illustrated.

\section{Model and numerical method}
 
In two dimensions, we consider   mixtures  
  of  anisotropic  and circular  
 particles with   numbers  $N_1$ and $N_2$, 
where   $N=N_1+N_2=4096$. 
The  concentration  of the circular  species is  
$c=N_2/N.$   The particle   positions are  
written as $\bi{r}_i$ ($i=1,\cdots,N$). 
The orientation vectors of the anisotropic  particles  
 are  expressed as 
$\bi{n}_i=(\cos\theta_i,\sin\theta_i)$
   in terms of angles $\theta_i$  ($i=1,\cdots,N_1$).     
The pair potential $U_{ij}$  between particles 
$i\in\alpha$ and $j\in\beta$ 
($\alpha,\beta=1,2$) is a  modified Lennard-Jones potential given by \cite{EPL} \be  
U_{ij}=4\ep\bigg[(1+ A_{ij}) 
\frac{\sigma^{12}_{\alpha\beta}}{r_{ij}^{12}}
-(1+ B_{ij})\frac{\sigma_{\alpha\beta}^6}{r_{ij}^6} \bigg], 
\en 
where $r_{ij}$ is the particle  distance 
and $\ep$ is the interaction energy. 
In terms of   characteristic lengths  
$\sigma_{1}$ and $\sigma_2$, we  set 
$\sigma_{\alpha\beta}=(\sigma_\alpha+\sigma_\beta)/2$. 
The potential is truncated  at $r_{ij}= 3\sigma_1$. 
The particle anisotropy 
is  accounted for  by the anisotropic 
factors $A_{ij}$ and $B_{ij}$, which    depend   on 
  the  angles  between   $\bi{n}_i$,  $\bi{n}_j$, and  
the relative direction 
$\hrij=r_{ij}^{-1}(\bi{r}_i-\bi{r}_j)$. In this 
paper, we  set  
\bea
A_{ij} &=&  \chi[ \delta_{\alpha 1} 
({\bi n}_i\cdot\hrij)^2+ 
\delta_{\beta 1} ({\bi n}_j\cdot\hrij)^2], \\
B_{ij} &=& \zeta [\delta_{\alpha 1}\delta_{\beta 2}
({\bi n}_i\cdot\hrij)^2+
 \delta_{\alpha 2}\delta_{\beta 1}
 ({\bi n}_j\cdot\hrij)^2], 
\ena
where  $\delta_{\alpha\beta}$ 
is the Kronecker delta,   
 $\chi$ is  the  anisotropy  strength of  
repulsion,  and  $\zeta$ is that of  attraction  between the two species.

The Newton equations  for ${\bi r}_i(t)$ and $\theta_i(t)$ 
 are given by  
\bea 
&&{m}\frac{d^2}{dt^2} {\bi{r}}_i=-\frac{\p{U}}{\p{\bi{r}_i}},\\
&&{I}\frac{d^2}{dt^2} {{\theta}}_i 
=- \frac{\p{U}}{\p{\theta_i}},
\ena
 where    
  $U=\sum_{i<j}U_{ij}$ is  the  total potential,  
$m$ is  the  mass common to  the two species,  
 and $I$ is the  moment of inertia. Note that Eq.(5) holds for 
the first species.
The total kinetic energy is given by 
$K=\sum_i m |d{\bi r}_i/dt|^2/2+ \sum_{i\le N_1} 
I |d\theta_i/dt|^2/2$.  Here,   $d\theta_i/dt$ 
is continuous in time 
and $\theta_i$ is  unbounded.

We  regard the anisotropic  particles   as ellipses. 
For  two anisotropic 
particles  $i$ and $j$,  
$U_{ij}$  is minimized at 
 $r_{ij}=2^{1/6}(1+A_{ij})^{1/6}\sigma_1$ 
as a function of $r_{ij}$ for fixed orientations. 
Then $A_{ij}$ is minimum  for ${\bi n}_i$ and 
$ {\bi n}_j $ being  perpendicular to 
${\hat{\bi r}}_{ij}$, 
while   it is maximum 
for ${\bi n}_i$ and $ {\bi n}_j$ 
being parallel to $\pm  {\hat{\bi r}}_{ij}$.
The  shortest and longest diameters are given by   
\be 
a_s=2^{1/6}\sigma_{1}, \quad a_\ell=(1+2\chi)^{1/6}2^{1/6}\sigma_{1}.
\en  
The  aspect ratio   is thus 
\be 
a_\ell/a_s=(1+2\chi)^{1/6}.
\en  
These  ellipses have  the 
area   $S_1=\pi{a_s}a_\ell/4$ and the  momentum of inertia  
$I=(a_\ell^2+a_s^2)m_1/16.$

In this paper, we fixed  the average 
 packing fraction    
$(S_1N_1+S_2N_2)/L^2$ at $0.95$,
 where $S_2=\pi2^{1/3}\sigma_{2}^2/4$. 
The cell length $L$ is  
about $70\sigma_1$.    We measure space in units of $\sigma_1$ 
and time  in  units of    
\be 
\tau_0=\sigma_1\sqrt{m/\ep}.
\en  
The  temperature is in units of 
$\epsilon/k_B$, 
where  $k_B$ is   the Boltzmann constant. 
In this paper, assuming small circular impurities, we set 
\be 
\sigma_2/\sigma_1=0.6, \quad 
\chi=1.2, \quad c=0.2.
\en 
The aspect ratio is then  $a_\ell/a_s= 1.23$ from Eq.(7),   
which is rather close to unity.  
If the aspect ratio is considerably larger than unity, 
liquid crystal order appears at 
higher  temperatures than in this paper.

We integrated the Newton equations 
using the leap-frog method 
under the periodic boundary 
condition. We lowered $T$  from 1 to 0.1 
at  a  cooling rate of  $dT/dt=0.9\times10^{-5}$.
We then changed $T$ to a final temperature 
and waited for $2\times10^5$, where  
    a Nos\'e-Hoover thermostat \cite{nose} was
imposed. But  after this initial preparation, 
we switched off the thermostat,  so our simulations 
have been performed in the  $NVE$ ensemble, where  
the average translational 
 kinetic energy was kept at $k_BT$ per particle.

Previously,  angle-dependent  potentials 
were  used for  liquid crystals \cite{Gay,An},  
water \cite{water2D}, 
 glass-forming liquids  \cite{Shintani}, and 
 lipids \cite{Leibler}. 
Our mixture system  is  similar  to that of 
 prolate Gay-Berne particles 
\cite{Gay}  and Lennard-Jones spheres studied 
by Antypov and  Cleaver \cite{An}.

\begin{figure}
\begin{center}
\includegraphics[width=240pt]{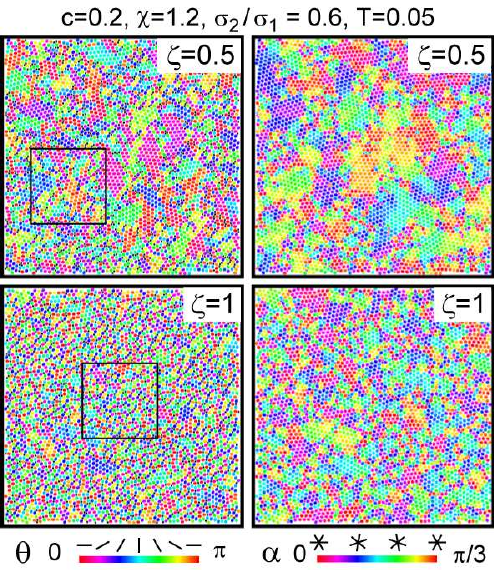}
\caption{(Color online) Orientational angles 
$\theta_j$ (left) and sixfold bond orientation angles 
$\alpha_j$ (right) in Eq.(10) for small  impurities 
with   $\zeta=0.5$ (top) 
and $1$ (bottom) at $T=0.05$  in double glass. 
Heterogeneities become finer with increasing $\zeta$.}
\end{center}
\end{figure}

\begin{figure}
\begin{center}
\includegraphics[width=240pt]{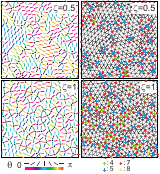}
\caption{(Color online) 
Left: Expanded snapshots of orientational angles $\theta_j$ 
around small impurities  in the box regions in the left panels of Fig.1.
Anchoring is homeotropic and  
impurity clustering is suppressed with increasing $\zeta$. 
Right: Delaunay diagrams, where marked 
are the particles with surrounding triangles different from six ($k\neq 6$).   
Those with $k=7$ and $8$ are mostly ellipses, 
while those with $k=4$ and $5$   
are mostly impurities. Here, $70\%$ impurities 
have  $k=5$ (fivefold  anchoring). }
\end{center}
\end{figure}

\begin{figure}[htbp]
\begin{center}
\includegraphics[width=240pt]{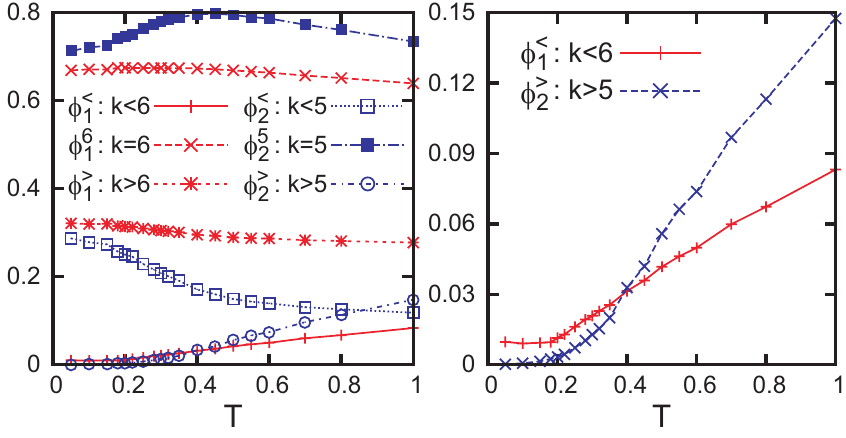}
\caption{(Color online) 
Left: Fractions  
$\phi_1^>$, $\phi_1^6$, 
and $\phi_1^<$  of  the ellipses 
with $k>6$,  $k=6$, and  $k<6$, respectively, 
and those  $\phi_2^>$,  $\phi_2^5$,  
and $\phi_2^<$  of the impurities 
with $k>5$,  $k=5$, and  $k<5$, respectively, 
as functions of $T$ with $\zeta=1$,  
where  $k$ is the number of the surrounding triangles in the 
Delaunay diagrams. 
Here,  $\phi_{1}^6$, $\phi_{1}^>$, $\phi_{2}^<$, and $\phi_{2}^5$  
are large at  any $T$,  but   $\phi_{1}^<$ and $\phi_{2}^>$  
decrease at low $T$. 
  Right: Fractions $\phi_1^<$ and $\phi_2^>$  vs $T$, 
  which are the fractions of liquidlike  defects \cite{Pro} 
 decreasing   at low $T$. 
}
\end{center}
\end{figure}

\section{Numerical results }
If $c\ll 1$,  our system   forms  
 an orientationally disordered crystal  
 (plastic solid)  below 
 a certain  $T$. It  then undergoes 
an orientational  phase transition with further lowering $T$.  
In this paper, we add   small isotropic 
 impurities as specified   in Eq.(9). 
Since  the size ratio $\sigma_2/\sigma_1$ is 
rather   small,  the positions can   be  highly   disordered 
as well as the  orientations, resulting 
in double glass at low $T$.  If the size ratio 
is close to unity, we obtain orientational glass at low 
$T$  with increasing $c$\cite{EPL}.

\subsection{Orientational  and positional configurations} 

 In  Fig.1, we  display 
snapshots of the particle angles and positions   
 at  $T=0.05$,  where 
the thermal fluctuations are very weak. 
In the left, we do not distinguish $\theta_j$ and $\theta_j\pm \pi$, 
so depicted are  $[\theta_j]_\pi =\theta_j- p\pi$ 
in the range $[0,\pi]$ with  an integer $p$. 
In the right, depicted are the  sixfold bond  angles $\alpha_j$  
 in the range $[0,\pi/3]$  defined  by   \cite{Nelson,Hamanaka} 
\be 
\sum_{k\in\textrm{\scriptsize{bonded}}} \exp [6i\theta_{jk}]
\propto  \exp[{6i\alpha_j}], 
\en 
where    $\theta_{jk}$ 
is the angle between ${\bi r}_{kj}= 
\bi{r}_k-\bi{r}_j$ and  the (horizontal) $x$ axis, 
 the summation is over 
the  particles $k$   within the range 
 $|\bi{r}_{jk}|<1.5\sigma_{\alpha\beta}$ (bonded to $j$), 
and $6\alpha_j$ is the   phase angle   of the left hand side. 
For $\zeta=0.5$ we can see  small 
orientationally ordered domains in the left 
and small polycrystal grains in the right.  
For  $\zeta=1$ both the orientations   and positions    
are more disordered, resulting in smaller   domains and grains. 
We remark   that 
 increasing the impurity concentration $c$ from zero 
 also gives rise to smaller  domains and grains \cite{Hamanaka,EPL}. 
Previously, similar  mesoscopic patterns 
of the orientations and the positions 
were  numerically  realized  in glassy states 
in the Shintani-Tanaka model  
\cite{Shintani}. 

The left panels of Fig.2 display  expanded snapshots 
of $\theta_j$ for $\zeta=0.5$ and 1, 
 where anchoring is 
homeotropic (perpendicular to the impurity surfaces) \cite{An}.  
Here, the impurity clustering is 
significant, which   took place during solidification 
\cite{EPL}.  However, with increasing  $\zeta$,  
the  impurities are  more strongly anchored 
 by  the surrounding ellipses  
and the aggregation of these {\it solvated} 
 impurities  is more    suppressed. 
Similar  homeotropic anchoring occurs  in water  around  
small ions as hydration due to the ion-dipole 
interaction \cite{Is}, which breaks   tetrahedral order 
resulting in   vitrification  at low $T$ 
\cite{water}.

In the right panels of Fig.2, 
we show the Delaunay triangulations  
of the  particle configurations 
in the left, which  are  the dual graphs  of  the Voronoi  diagrams. 
Here,   each particle is surrounded 
by several triangles,  so let $k$ be the number of these 
 triangles, which has the meaning of the coordination number. For a 
hexagonal lattice, we have $k=6$. Thus, in these panels, 
we mark  the  noncrystalline particles 
with $k\neq 6$,    where those with  $k=7$ or $8$ 
are mostly ellipses  and those  with   $k=4$ or   5 
are impurities.  For $c=0.2$, 
a  majority of the ellipses ($\sim 65\%$) 
have  $k=6$  in the presence of  a considerable fraction of 
 small  crystalline regions, 
while a majority of the impurities  ($\sim 70\%$) 
have $k=5$ due to the homeotropic 
anchoring of the surrounding  ellipses. 

In the left panel of Fig.3,  we  display the 
 fractions of  the ellipses  
with $k>6$, $k=6$, and $k<6$ and 
 those of the impurities  
with $k>5$, $k=5$, and $k<5$. 
These six fractions are denoted by 
$\phi_1^>$, $\phi_1^6$,  $\phi_1^<$, 
 $\phi_2^>$, $\phi_2^5$, and $\phi_2^<$, respectively, 
 as functions of $T$.   At low $T$,  
  $k$ is mostly  $6$ or $7$ for the ellipses 
and is mostly $4$ or $ 5$ for the impurities. In fact, 
for the data in Figs.1 and 2 at $T=0.05$, 
we have $(\phi_{1}^6, \phi_{1}^>) 
\cong (0.66,0.33)$ and $(\phi_{2}^<, \phi_{2}^5) 
\cong (0.28,0.71)$ for 
$\zeta=1$, while these sets are  
$ (0.73, 0.26)$ and $(0.05, 0.94)$, respectively,  for $\zeta=0.5$. 
In the right panel of Fig.3, 
$\phi_1^<$ and $\phi_2^>$  are  very small at 
low $T$ and  increase with increasing $T$. 
Thus, the ellipses  with  $k <  6$ and 
the impurities with  $k > 5$ represent   
liquidlike defects  \cite{Pro}.

  Hentschel {\it et al.} \cite{Pro} 
studied the positional disorder 
using the  Voronoi graphs   for a mixture of 
circular particles with the soft-core potential in two dimensions.
In their simulation,   small (large) particles enclosed by heptagons 
(pentagons) form liquidlike defects  decreasing   at low $T$. 

\subsection{Time-correlation functions}

\begin{figure} 
\begin{center}
\includegraphics[width=200pt]{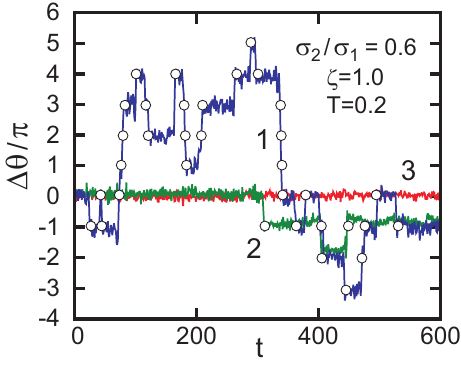}
\caption{(Color online)
 Time-evolution of 
angle changes  $\Delta\theta_i(t_0,t_0+t)$ 
in Eq.(11)  for (1) a frequently flipping  ellipse, 
(2) an infrequently  flipping one, and (3) an 
inactive one  for $\zeta=1$ and $T=0.1$. 
Flip events occur at points ($\circ$) 
on the curves (see the appendix). 
These jumps are very different from thermal vibrations 
but occur thermally. 
}
\end{center}
\end{figure}

\begin{figure} 
\begin{center}
\includegraphics[width=190pt]{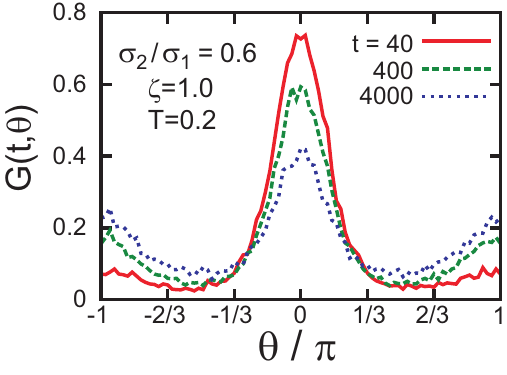}
\caption{(Color online) 
Time-dependent angle distribution function 
$G(t,\theta)$ in Eq.(12) at $t=40, 400$, and $4000$ 
for $\zeta=1$ and $T=0.2$. 
Peaks emerge at $\theta = \pm \pi$ due to flip motions 
on the time scale of $\tau_1=400$. Afterwards, 
 $G(t,\theta)\to 1/2\pi$ on the 
time scale of $\tau_2=24000$.  }
\end{center}
\end{figure}

\begin{figure} 
\begin{center}
\includegraphics[width=230pt]{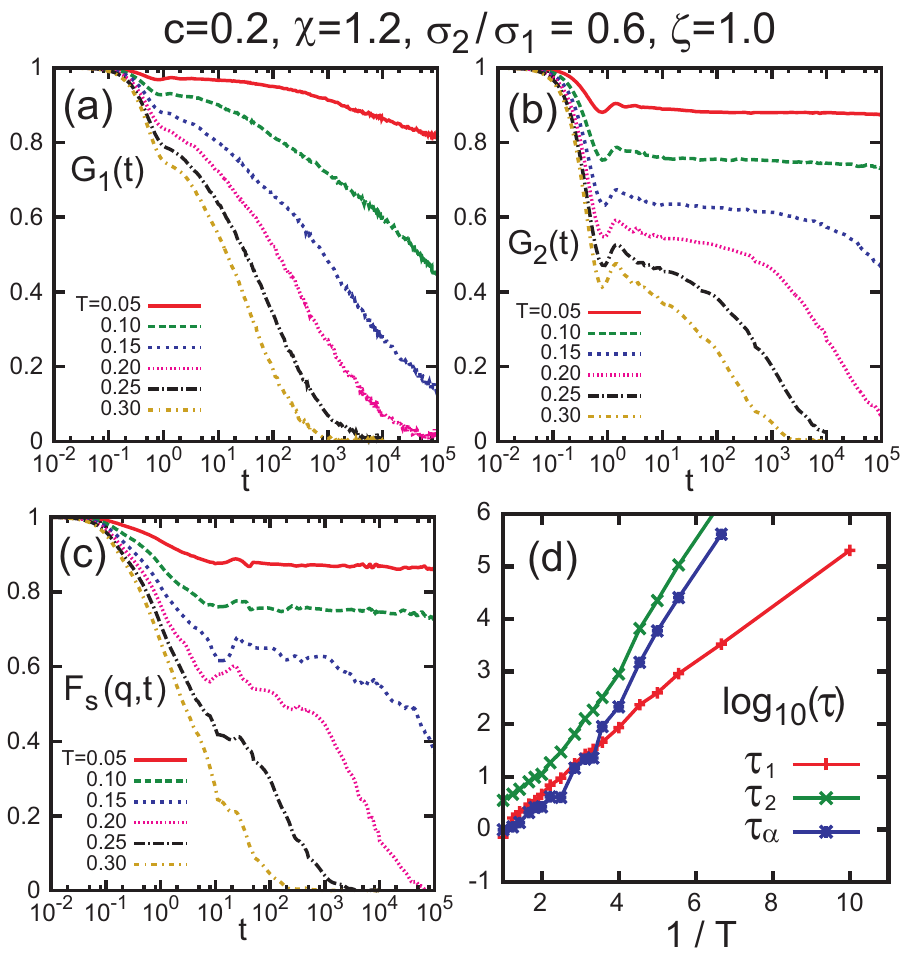}
\caption{ (Color online) 
(a) $G_1(t)$,   (b) $G_2(t)$, 
 (c) $F_{s}(q,t)$   at $q=2\pi$ for ellipses at six $T$. 
 (d) Relaxation times  $\tau_1$, $\tau_2$, and $\tau_{\alpha}$ 
in Eqs.(14)-(16) vs $1/T$. 
Here,  $\zeta=1$ and time $t$ is  in units of 
$\tau_0$ in Eq.(8).}
\end{center}
\end{figure}

\begin{figure}[htbp]
\begin{center}
\includegraphics[width=240pt]{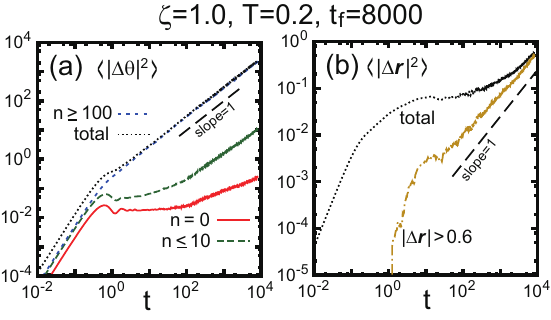}
\caption{(Color online)
 Angular and positional mean-square displacements 
for $\zeta=1$ and $T=0.2$.  
(a)  Angular one   $M_\theta(t)$ in Eq.(17) and 
contributions from those with $n_i\ge 100$, $n_i\le 10$, and 
$n_i=0$, where $n_i$ is the flip number  
for  $t_f=20\tau_1=8000$ (see the appendix). The contribution from 
$n_i \ge 100$ approaches $M_\theta(t)$ for $t \gs 1$, 
leading to $ D_R=0.14$. 
(b) Positional  one  $M (t)$  in Eq.(18) and 
contribution from those with 
$\Delta { r}_i  >0.6$  in Eq.(21), where the latter 
 grows  linearly  for $t\gs 20$ with  
$D=1.4 \times  10^{-5}$. 
}
\end{center}
\end{figure}

For  strong short-range anchoring,  
the rotational dynamics 
sensitively depends on whether 
the anisotropic particles  are close or  
far from the impurities. 
In Fig.4, we show  time-evolution   of  the angle changes,  
\be 
\Delta\theta_i(t_0,t+t_0) = 
\theta_i(t+t_0)-\theta_i(t_0),  
\en  
where  we pick up a rapidly rotating ellipse, 
a rarely  flipping one, and an inactive one. 
We can see instantaneous flip motions 
by $\pm \pi$.  In  previous simulations 
in glassy states,  they observed 
flips  for  rod-like molecules 
\cite{Kob1,Klein,Michele} and  large angle jumps  for 
 ortho-terphenyl (OTP)   \cite{Lewis,Debe}.

We  introduce   the  distribution  of the angle changes, 
\be 
G(t,\theta)
=\frac{1}{N_1}\sum_{ i \in 1}
 \av{\delta([\Delta\theta_i(t)]_{2\pi} -\theta)} ,
\en 
where $\Delta\theta_i(t)= 
\Delta\theta_i(t_0,t+t_0)$  and  
$-\pi \le \theta<\pi$. For any angle $\varphi$, we define  
  $[\varphi]_{2\pi}=\varphi -2p\pi$, which is  
in the range $[-\pi,\pi]$ with an integer $p$.  
 Furthermore, we consider 
the $\ell$-th  moments of $G(t,\theta)$ 
given by  
\bea 
G_\ell(t)&=& \int_{-\pi}^{\pi}d\theta G(t,\theta) \cos(\ell\theta)\nonumber\\
&=& \frac{1}{N_1}\sum_{ i \in 1}
 \av{\cos[\ell\Delta\theta_i(t)]} .
\ena   
We calculated $G(t,\theta)$, $G_1(t)$, and $G_2(t)$ 
by taking  the  average $\av{\cdots}$  over 
the initial time $t_0$ and over five  runs.

In Fig.5, we show time-evolution of 
$G(t,\theta)$, 
where  flip motions give 
rise to peaks  at $\theta = \pm\pi$ growing on the time scale of $\tau_1$. 
Thus, these flip motions  cause the decay 
of  $G_1(t)$ in Fig.6(a).  
However, $G_2(t)$ is unchanged by the turnovers and 
 decays more slowly after the initial relaxation    in Fig.6(b).  
Notice that  $G(t,\theta)$ 
tends to $1/2\pi$ on  the time scale of   $\tau_2$. 
In Fig.6(c), we also show  
 the self part of the density time-correlation 
function $F_{s}(q,t)$ at $q=2\pi$ for the 
ellipses, which closely  resembles  $G_2(t)$.

We define the   relaxation times 
$\tau_{1}$,  $\tau_{2}$, and 
${\tau_{\alpha}}$   as 
\bea 
&&G_1(\tau_{1})=1/e, \\ 
&& G_2(t)\propto\exp[-(t/\tau_{2})^{\beta}] \quad \quad (t>1),\\
&& F_{s}(q,t)\propto\exp[-(t/{\tau_{\alpha}})^{\gamma}]~\quad (t>1),  
\ena 
where the  exponents $\beta$ and $\gamma$  are about 0.4  for $T\ls 0.2$. 
We here determine  $\tau_{2}$ and 
${\tau_{\alpha}}$ from the long-time 
relaxations of  $G_2(t)$ and  $F_{s}(q,t)$, respectively. 
In Fig.6(d), we plot them, where 
 $\tau_{1}\ll   {\tau_\alpha} \sim \tau_{2}$. 
In all the $T$ range  in Fig.6(d), 
 $\tau_{1}$ may be nicely  fitted to   
the  Arrhenius form, $\ln(\tau_1)=1.4/T-1.2$.   
On the other hand, $\tau_{2}$ and $  {\tau_\alpha}$ 
exhibit a changeover at $T \sim 0.2$ and 
can be   fitted to   the  Arrhenius forms as 
$\ln(\tau_2)=2.6/T-3.0$ and $\ln(\tau_\alpha)=2.6/T-4.5$, for $T\ls  0.2$, 
so   $\tau_2/\tau_\alpha\cong 4$ for $T\ls  0.2$.  
Thus,   $G_1(t)$ decays mainly due to  thermally activated  
 flip motions, which  are nearly 
 decoupled from the translational motions. 
On the other hand,  $G_2(t)$ and $F_s(q,t)$ 
  decay at longer times due to  irreversible 
 configuration changes involving at least several particles.

In three dimensions, the distribution of 
 the angles $\cos^{-1}[{\bi n}_i(t_0+t)\cdot{\bi n}_i (t_0)]$   
was calculated 
 for OTP \cite{Lewis} and  for dumbbells \cite{Kob1,Michele}. 
In these papers,   this  distribution 
 exhibited peaks due to orientational jumps. 
Also in  the rotational 
time-correlation functions  in three dimensions, 
the Legendre polynomials $P_\ell(u_i(t))$ 
with $u_i(t) ={\bi n}_i(t_0+t)\cdot{\bi n}_i (t_0)$ 
were  used   
\cite{Klein,Michele,Kob1,Chong,Chong1,Moreno,Debe}, 
where $G_\ell(t)$ with even $\ell$ 
decayed slower than those with odd $\ell$ at low $T$. 
These previous findings are in accord with our results.

\subsection{Mean-square displacements}

In the literature, 
the  angular  mean-square  displacement has been 
calculated to study  the rotational diffusion 
\cite{Michele,Chong,Chong1,Debe,Kob1}. 
In two dimensions, it is defined by    
\be 
M_\theta(t)= \av{|\Delta\theta|^2}= \frac{1}{N_1} 
\sum_{i \in 1}  \av{|\Delta\theta_i(t_0,t_0+t)|^2}.
\en 
We also introduce   the usual positional 
mean-square displacement for the ellipses by  
\be 
M(t)= \av{|\Delta{\bi r}|^2}= \frac{1}{N_1} 
\sum_{i\in 1} \av{ |\Delta{\bi r}_i(t_0,t_0+t)|^2}, 
\en 
where $\Delta{\bi r}_i(t_0,t_0+t)={\bi r}_i(t_0+t)- {\bi r}_i(t_0)$. 
At very short times,  these quantities 
exhibit the ballistic behavior $(\propto t^2$). 
At long times, they  grow linearly 
in time as  
\bea 
&&M_\theta(t) \cong 2D_Rt ,\\
&& M(t) \cong 4D t ,
\ena 
where $D_R$ and $D$ are  the rotational and translational 
diffusion constants, respectively. 
If the rotational activity 
is  strongly  heterogeneous, 
 the so-called 
Stokes-Einstein-Debye relation 
$D_R \sim  k_BT/\eta a$  
does not hold \cite{Silescu,Debe}, where $\eta$ is 
the viscosity and $a$  is the radius of the diffusing 
particle.  In our case, 
$M_\theta(t)$  is greatly increased 
by  rapidly flipping  ellipses and  
$D_R$ from it does not correspond to 
any  experimentally observed 
relaxation times at low $T$.

%

In  Fig.7(a),  we plot 
 $M_\theta(t)$  for $\zeta=1$ and $T=0.2$.
Here,   it   attains the diffusion behavior 
with $D_R=0.14$   for $t\gs 1$, while $G_1(t)$ 
decays slower with   $\tau_1= 400$. 
We also display  the contributions  
to the sum in $M_\theta(t)$ 
in Eq.(17) from the ellipses with 
$n_i=0$, $n_i\le 10$, and $n_i \ge 100$,
where $n_i$ is the flip number 
of ellipse $i$ in a time interval with width  
$t_f=8000$ (see the appendix).  The fractions 
of these three groups are $0.16$, $0.34$, 
and $0.39$, respectively.  Remarkably, 
 the contribution from $n_i \ge 100$ 
 approaches $M_\theta(t)$ for $t\gs 1$, while 
that from $n_i \le 10$ 
 behaves diffusively as  $0.7\times 10^{-3} 
\times 2t$ for $t\gs \tau_1$.   
Thus, the effective rotational diffusion constant 
of the ellipses with $n\le 10$ 
is $0.7\times 10^{-3}/0.34= 2\times 10^{-3}$. 
The $M_\theta(t)$ itself exhibits the plateau behavior 
at much lower temperatures (say, $T=0.05$), 
while the contributions 
from $n=0$ and $n\le 10$ exhibit it at $T=0.2$.  

In Fig.7(b),   for $\zeta=1$ and $T=0.2$, 
$M(t)$ still 
in the course of plateau-to-diffusion 
crossover even at  $t=10^4$. 
To obtain small  $D$, we also plot 
the contribution from the ellipses 
with large displacements \cite{Kawasaki-OnukiPRE}, 
\be 
M^>(t)= \frac{1}{N_1} \sum_{i\in 1}\av{ 
\Theta(\Delta{r}_i(t) -\ell_c) |\Delta{r}_i(t)|^2},
\en 
where  $\Delta r_i (t)$ 
is an abbreviation of 
$|\Delta{\bi r}_i(t_0,t_0+t)|$ and 
$\Theta(u)$ is the step function 
being equal to 1 for $u\ge 0$ and to 0 for $u<0$.  
The threshold length  $\ell_c$ is set equal to $ 0.6$. 
In this  restricted   sum,  
the thermal vibrational motions within transient cages 
are excluded, so  it picks up the 
thermally activated jumps only. As a result,   we have  
 the linear growth   $M^>(t)\cong 4Dt$  
with $D=1.4 \times  10^{-5}
\sim 10^{-4}D_R$ from the early stage $t\gs 20$. 
  This behavior of $M^>(t)$ is insensitive to a small 
change of $\ell_c$ \cite{Kawasaki-OnukiPRE}. 
For example, almost the same results 
followed for  $\ell_c=0.8$. 

As a similar observation, Chong and Kob found that 
$D_R \tau_2$ grows strongly 
with lowering $T$ for a mixture of 
rigid  dumbbell molecules \cite{Chong}.
See more discussions 
for other molecular systems in the 
item (i) in   the summary.

\subsection{Distribution of flip numbers 
 }

\begin{figure}
\begin{center}
\includegraphics[width=250pt]{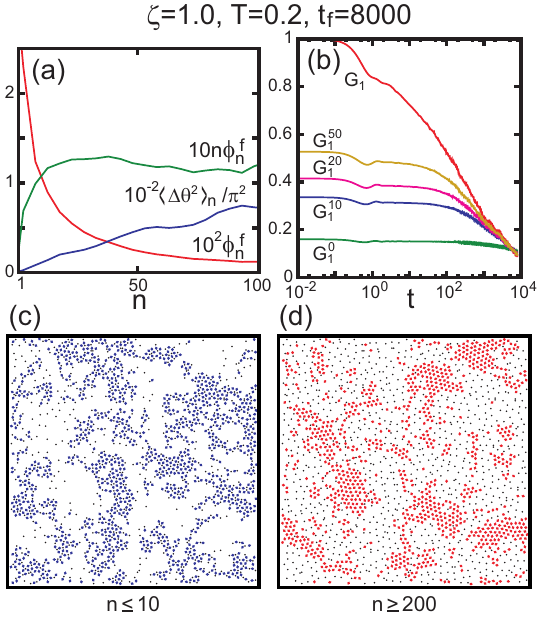}
\caption{(Color online) (a) $10^2\phi_n^f$, 
$10 n\phi_n^f$  for $n\ge 1$, and 
$10^{-2}\av{|\Delta\theta|^2}_n/\pi^2 $ in Eq.(23), 
where $t_f=20\tau_1=8000$ (averages over six runs). 
   (b) $G_1(t)$ in Eq.(13) and $G_1^n(t)$ in Eq.(28) 
with $n=0, 10, 20,$ and 50  
at $t=t_f$. where the latter approach 
the former at long times. 
Snapshots of ellipses   
with $n_i \le 10$ in (c) and  those with 
$n_i\ge 200$ in (d), whose heterogeneities are 
correlated with the impurity clustering. }
\end{center}
\end{figure}

\begin{figure}
\begin{center}
\includegraphics[width=250pt]{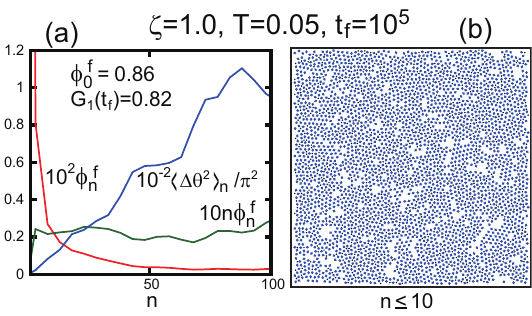}
\caption{ (Color online)  $10^2\phi_n^f$, 
$10 n\phi_n^f$, and $10^{-2}
\av{|\Delta\theta|^2}_n /\pi^2$ (averages over six runs)
 in (a)  and snapshot of ellipses   
with $n\le 10$ in (b), where $\zeta=1$, $T=0.05$, and $t_f
=10^5$. At this low  $T$, $\phi_0^f=0.86$ and 
the fraction of the depicted ellipses 
in (b) is $0.92$. 
}
\end{center}
\end{figure}

The  large size of $D_R$ is due to the presence of 
ellipses frequently undergoing flip motions. 
In the appendix, we will give a method 
of determining the flip number $n_i$ 
for each ellipse $i$ in a  
time interval $[t_0,t_0+t_f]$.  
We should choose a sufficiently 
large width $t_f $ to detect a wide range 
of $n_i$. In  the following, 
  $t_f=8000$ at $T=0.2$ in Figs.7 and 8 
and  $t_f=10^5$ at $T=0.05$ in Fig.9.
The curves in Figs.8(a), 8(b), and 9(a) 
are the averages over six runs.

 For a given time interval $[t_0, t_0+t_f]$,  
the   fraction of  the ellipses 
with $n$ flips is written as   
\be 
\phi_n^f = \sum_{i\in 1}\av{\delta_{nn_i}} /N_1. 
\en 
We further 
 introduce  the $n$-dependent 
mean-square displacement among the ellipses with 
$n$ flips as  
\be 
\av{|\Delta\theta|^2}_n (t) = 
\frac{1}{N_1\phi_n^f}  \sum_{i \in 1}\av{ 
 \delta_{n n_i} |\Delta\theta_i(t_0,t_0+t)|^2}.
\en 
It follows the sum relation, 
\be 
M_\theta(t) = \sum_{n \ge 0}\phi_n^f  
 \av{|\Delta\theta|^2}_n (t) .
\en 
In our case, most ellipses 
undergo  $+\pi$ flips and $-\pi$ flips equally 
on long times, so   $ \sum_{i \in 1}\av{ 
 \delta_{n n_i} \Delta\theta_i(t_0,t_0+t)}=0$.

In Fig.8(a), we plot    $\phi_n^f$, $n\phi_n^f$, 
and $\av{|\Delta\theta|^2}_n (t_f)$ (where $t=t_f$) for $\zeta=1$ and 
$T=0.2$.
Here, the fraction of the ellipses  with no flip  
is given by  $\phi_0^f=0.16$. We find that 
the flip number distribution is very broad as
\be 
\phi_n^f \sim 0.1 n^{-1}, 
\en 
in the range  $1 \ll  n<n_{\rm max}$, where 
 $n_{\rm max}$ is an  upper bound about $10^3$. 
In the present case,   
 the sums of $\phi_n^f$ in the ranges $1\le n\le 10$, 
 $11 \le n\le 99$,  and $ n\ge 100$ 
are  $ 0.18$,  $ 0.27$, and $ 0.39$, respectively. 
Furthermore, we  find   
\be 
\av{|\Delta\theta|^2}_n (t) \sim \pi^2 n_t = \pi^2 n t/t_f  
\en 
for $t\gg 1$ and $n\gg 1$.   In  a general  
time  width $t$,  the ellipses with $n$ flips 
in the reference time width $t_f$ 
should flip $n_t=  nt/t_f$ times  on the  average,  where 
$t\gg 1$ and $n\gg 1$. Then, together with 
the sentence below Eq.(24), 
 Eq.(26) is a natural relation. 
 From Eqs.(24)-(26)  we find 
\be 
D_R \sim  n_{\rm max} /t_f ,
\en 
which means that  $D_R$ is determined by  rapidly rotating ellipses. 
To be self-consistent, 
 $n_{\rm max}$ should be proportional to $t_f$; 
then,  $D_R$ is independent of  $t_f$. 

In Fig.8(b), we compare 
 $G_1(t)$ and  the  restricted  sums, 
 \be 
G_1^n(t) =  \frac{1}{N_1} \sum_{i \in 1 }\AV{ 
\Theta(n-n_i)
\cos[\Delta \theta_i(t_0, t_0+t)]}   , 
\en 
where we set  $n=0, 10, 20$, and $50$.  We here pick up the ellipses 
with flip numbers not exceeding $n$ owing to the step function $\Theta$. 
We can see that  these $G_1^n(t)$  are nearly 
constant for some time and become nearly equal to   
$G_1(t)$ after long times. For $t \gs 10^3$, 
$G_1(t)$ is composed of the contributions 
from the ellipses with $n \le 10$.

In Figs.8(c) and (d), we show snapshots of the ellipses 
with $n\le 10$ and $n \ge 200$, respectively. The distributions of these 
rotationally inactive and active  ellipses are highly heterogeneous. 
This marked feature is rather natural in view of 
the  mild aspect ratio $1.23$  and 
the significant impurity clustering. 
In fact, the impurities are nearly  absent  
in the red regions in Fig.8(b).

In Fig.9, we also show that the 
 flip motions still remain  
even at  $T=0.05$.  In this case, 
we find $G_1(t)\sim 0.8$ 
 at $t=10^5$ in Fig.6(a),  but we estimate  $\tau_1 \sim 10^{12}$ from 
the extrapolation of the Arrhenius form (see the sentences below Eq.(16)). 
In Fig.9(a), we find  $\phi_0^f=0.86$ and  
  $\phi_n^f\sim 0.02n^{-1}$ and again obtain Eq.(26) for 
  $\zeta=1$, $T=0.05$, and  $t_f=10^5$. 
In Fig.9(b),  displayed is  a  snapshot of the  ellipses   
with $n_i \le 10$, whose fraction is 
$\sum_{n \le 10} \phi_n^f=0.92$. 
 Even at this low $T$,  $2\%$ ellipses 
have $n_i>200$. We have  $D_R=1.2\times 10^{-3}$ 
due to these  rapidly flipping ellipses.
This snapshot was produced by the 
initial particle configuration  common 
to that in Fig.8(b). Most of  the ellipses 
in the red regions in Fig.8(b) 
 are   now inactive, 
since their orientation 
alignment increases 
with lowering $T$.




\begin{figure}[htbp]
\begin{center}
\includegraphics[width=240pt]{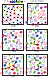}
\caption{ (Color online) Time-development of a configuration 
change at   successive 
times  $ t_0+t$  with $t=0,2,4, 6,8$, 
and $10$ for $\zeta=1$  and $T=0.2$. 
Ellipses  with  
$|\Delta {\bi r}_i (t_0,t_0+t)| >0.6$ 
or  $c_i=\cos(2|\Delta \theta(t_0,t_0+t)|)<0.2$ 
are written. Arrows represent 
$\Delta {\bi r}_i (t_0,t_0+t)$ and colors 
$ \theta_i(t_0+t)$  
according the color bar as in Fig.1. 
Impurities with $|\Delta {\bi r}_i (t_0,t_0+t)|>0.6$ 
are written as  black circles ($\bullet$). 
Other particles are written as   white ellipses or circles. 
}
\end{center}
\end{figure}

\begin{figure}[htbp]
\begin{center}
\includegraphics[width=240pt]{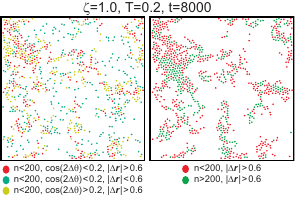}
\caption{(Color online) 
Comparison of ellipses  with  $c_i \equiv  
\cos[2\Delta{\theta}_i(t_0, t_0+t)]
<0.2$   and particles  with large  displacement 
$\Delta r_i \equiv |\Delta{\bi r}_i(t_0, t_0+t)|>0.6$ 
 for $\zeta=1$  and $T=0.2$ at $t=t_f= 8000$. 
In the left,  depicted  are 
three groups of ellipses 
with  $c_i <0.2$  and $\Delta { r}_i >0.6$ (in red), 
$c_i <0.2$  and $\Delta { r}_i <0.6$ (in green),
 $c_i >0.2$ and $\Delta { r}_i >0.6$ (in yellow), 
 whose 
fractions are 0.08, 0.10, and 0.09, respectively.  
In the right, depicted  are $0.18N_1$ ellipses 
with $\Delta { r}_i >0.6$  and $n_i <200$  (in red) 
 and $0.11N_1$  ones with $\Delta { r}_i >0.6$  and $n_i >200$  
(in green). }
\end{center}
\end{figure}

\subsection{Dynamic heterogeneities in 
 non-flip rotations and displacements}

Now, we examine 
the long-time structural relaxation 
caused by collective configuration changes, 
where  large  displacements induce 
 large non-flip rotations. 
These events should occur repeated 
in the same fragile regions 
 on  time scales longer than $\tau_2$  
\cite{Shiba,Kawasaki-OnukiPRE}.

In Fig.10, we illustrate 
 time-development of a configuration change 
at successive times  $t_0+t$  with
 $0\le t  \le 10$.  Depicted are the ellipses  with  
$\Delta r_i(t)= |\Delta {\bi r}_i (t_0,t_0+t)| >0.6$ 
or  $c_i (t)<0.2$, where  
\be 
c_i(t)= \cos[2\Delta{\theta}_i(t_0, t_0+t)].  
\en 
The condition $c_i(t)<0.2$ 
means  $0.22 \pi<|\Delta{\theta}_i(t_0, t_0+t)|< 0.78 \pi$  
in the range $[-\pi,\pi]$. 
From  Eq.(13) we have  
$G_2(t)= \sum_{i\in 1}\av{c_i(t)}/N_1$.
In the  narrow  region in Fig.10, the particle configuration 
was nearly stationary for $t \le 0$, 
but large particle motions  started for $t>0$ 
and continued on a time scale of 10. 
We can see  circulating particle motions at $t=6$ and 8 
and  stringlike ones at $t=10$ 
\cite{Donati,Glotzer,Shiba,Kawasaki-OnukiPRE}. 
The orientations of these ellipses are largely changing with 
their movements.  
For  $t>10$,  the subsequent displacements 
became   small, but  considerable  orientational 
motions  persisted until $t \sim 20$.

We examine the non-flip motions in terms of 
$c_i(t)$ in Eq.(29), since it   is invariant with respect to turnovers.  
In Fig.11(a),  we  visualize  the 
correlation between $c_i= c_i(t)$ and   $\Delta r_i= \Delta r_i(t)
$ for $\zeta=1$ and  $ T=0.2$. We set   $t=t_f=8000$, 
which  is one-third of $\tau_2 \sim 24000 $ ($\sim 4 \tau_\alpha)$. 
We present a snapshot of 
the ellipses with 
(a)  $c_i <0.2$ and $\Delta { r}_i > 0.6$ ,
(b) $c_i <0.2$ and $\Delta { r}_i<0.6$,  
and (c) $c_i >0.2$ and  $\Delta { r}_i >0.6$.
  Here, we exclude the ellipses with $n_i >200$ 
$(\sim 30\%)$, because they do not exhibit the glassy behavior. 
 The fractions of these depicted groups 
are (a) 0.08, (b) 0.10, and (c) 0.09, 
while the fraction of the ellipses with 
$c_i >0.2$,  $\Delta { r}_i <0.6$, and $n_i\le 200$ is $0.41$.
Thus, if we consider the ellipses with $c_i<0.2$ and $n_i<200$, 
a half of them have undergone displacements with $\Delta r_i >0.6$. 
Also, if we consider the ellipses with $\Delta r_i >0.6$  and $n_i<200$, 
a half of them have undergone large angle changes 
with  $c_i <0.2$. 
  
In  Fig.11(b),  displayed is  a  snapshot  of 
the ellipses with (a) $\Delta { r}_i >0.6$ and $n_i<200$ ($18\%$) 
and (b) $\Delta { r}_i >0.6$ and $n_i>200$  ($11\%$).
The group (a) here consists of the groups (a) and (b) in Fig.11(a). 
Here,  the ellipses in these two groups form 
clusters, indicating collective displacements. 
In addition, clusters of one group 
are adjacent to those of another group.
 Thus, rotationally active ellipses with large  $n_i$ 
tend to be translationally active also.

\section{Summary and remarks}

 We have performed 
simulation of  a mixture of elliptic 
particle with a mild aspect ratio ($=1.23$) 
and smaller circular impurities with $\sigma_2/\sigma_1=0.6$ 
at $20\%$. 
We have assumed   an angle-dependent 
attractive interaction between the ellipses 
and the impurities ($\propto \zeta$), which leads  to 
 the homeotropic anchoring 
of the ellipses around each impurity. 
We summarize our main simulation results.\\
1) We have shown snapshots of 
the orientations and the positions in Figs.1 and 2, which 
are  mesoscopically  heterogeneous. 
From the Delaunay triangulation in the right panels of Fig.2, 
we have found that  the number of surrounding triangles 
(the coordination number) is 6 or 7 for the ellipses 
and 5 or 4 for the impurities in glassy states, as plotted  in Fig.3. 
A majority of the impurities ($\sim 70\%$) are surrounded by 
5 ellipses, analogously to the case of 
the Shintani-Tanaka model\cite{Shintani}.\\
2) We have calculated the distribution function 
of the angle changes $G(t,\theta)$ in Eq.(12), 
which  exhibits 
 peaks at $\theta =\pm\pi$ for large $t$ due to flip motions as 
in Fig.5. We have found that 
the rotational time-correlation functions $G_1(t)$ and $G_2(t)$ 
of the ellipses 
relax very differently at long times in Fig.6, because 
$G_1(t)$ decays due to flip motions and $G_2(t) $ 
due to configuration changes.\\   
3)  We have found that 
the angular mean-square displacement 
$M_\theta(t)$  in Eq.(17) behaves as  
$2D_Rt$ rapidly for $t\gs 1$ with very large $D_R$ in Fig.7. 
This is in marked contrast to  
the  slow time-evolution of 
 the translational  mean-square displacement 
$M(t)$.  However, the contribution to 
$M(t)$ from the largely displaced ellipses ($|\Delta r_i|>0.6$) 
has exhibited  the diffusion behavior with $D= 10^{-4}D_R$, 
because the diffusion is governed by  the activation dynamics
\cite{Kawasaki-OnukiPRE}.
\\
4) We have displayed  the fractions  $\phi_n^f$ 
of the ellipses with $n$ flips 
in a time interval $[t_0,t_0+t_f]$, 
where  $t_f=8000$ at $T=0.2$ in Figs.7 and 8 
and $t_f=10^5$ at $T=0.05$ in Fig.9. We have found 
a very broad  distribution  $\phi_n^f (\propto  n^{-1})$ 
for $1\ll n<n_{\rm max}$. The   angular mean-square displacement 
$\av{|\Delta\theta|^2}_n(t)$  among the ellipses 
with $n$ flips behaves as $\pi^2nt/t_f$. 
Then $D_R \sim  n_{\rm max}/t_f$ due to 
 rapidly flipping ellipses. We have also shown that 
 the long-time decay of $G_1(t)$ is  determined by 
ellipses with $n \le 10$ in Fig.8(b).\\
5)  
The distributions of the rotationally inactive ($n\le 10$) 
and active ($n\ge 200$) ellipses at  $T=0.2$ have been 
presented in Figs.8(c) and (d), while that of the inactive ones 
at $T=0.05$ has been given in Fig.9(b).\\
6) 
We have illustrated  time-development of a 
configuration change in Fig.10, 
where large displacements and  non-flip 
rotations are coupled. 
We have demonstrated 
close correlation between  
 non-flip rotations and large displacements in Fig.11.\\

We make  remarks as follows.\\
(i) Our  potential energy is invariant with 
respect to   turnovers. This is also the case of 
diatomic molecules or dumbbells \cite{Chong,Moreno}, 
for which  the flip motions should be highly heterogeneous 
  for mild aspect ratios.   For methanol,   
Sindzingre and Klein \cite{Klein} 
found  flip motions near the 
glass transition. For OTP, 
 Lewis and  Wahnstr$\ddot{\rm o}$m \cite{Lewis} 
found  translation-free orientational jumps, while   
Lombardo {\it et al.}  \cite{Debe}  found 
an enhancement in the rotational 
motions  relative to the 
translation motions at low $T$. 
For these systems, the heterogeneity  
of  orientational jumps should be 
examined in more detail.  
\\
(ii) We have chosen a mild aspect ratio ($=1.23)$ 
to find significant flip motions. However, 
an  increase  in  the aspect ratio  leads to 
 a decrease in  the flip frequency,  
on which we will report shortly. \\  
(iii) 
We have suppressed the clustering of 
small impurities by the angle-dependent 
attractive interaction. 
If we consider large  impurities (say, $\sigma_2/\sigma_1=1.4$), 
we may realize double glass by adding  a repulsive interaction 
among the impurities  suppressing  crystal formation.\\ 
(iv) 
The phase behavior of 
mixtures of two species of anisotropic particles 
should be studied in future, where we expect  
 nematic or smectic  glass.\\  
(v) 
 The spatial scales of the structural 
 heterogeneities   depend on various parameters. 
If the  oriented domains are not too small,   
there arises a  large orientation-strain coupling, 
leading to   soft elasticity and 
 a shape-memory effect \cite{EPL}. These  effects 
were observed  for  Ti-Ni  alloys \cite{Ren}
(where atomic displacements 
 within unit cells cause structural   changes). 
When anisotropic particles  have electric dipoles\cite{ori}, 
 mesoscopic  polar  domains  appear     
as in ferroelectric glass (relaxors) \cite{Cowley}.
Including such  metallic  alloys also, 
we point out relevance of the compositional 
heterogeneity  in the development 
of  mesoscopic order. 
\\

\begin{acknowledgments}
This work was supported by Grant-in-Aid 
for Scientific Research  from the Ministry of Education, 
Culture,  Sports, Science and Technology of Japan. 
K. T. was supported by the Japan Society for Promotion of Science.
The present numerical calculations were carried out on 
SR16000 at YITP in Kyoto University.  
\end{acknowledgments}

\vspace{2mm}
\noindent{\bf Appendix: 
Flip events in numerical analysis}\\
\setcounter{equation}{0}
\renewcommand{\theequation}{A\arabic{equation}}

In our numerical analysis, we determine  a series of   
flip times, $t_{i1},t_{i2},t_{i3},\cdots$   
for each ellipse $i$. We write   
the angle change  as  $\Delta\theta_i (t)
=\Delta\theta_i (t_0,t+t_0)$, suppressing $t_0$.  
(i) At the first flip time $t_{i1}$ 
we set 
\be 
|\Delta\theta_i(t_{i1}) |= 2\pi/3.
\en  
  For $t>t_{i1}$ we introduce  
\be 
\Delta\theta_{i1} (t) = 
\Delta\theta_i(t) \pm \pi,
\en 
 where $+\pi$ or $-\pi$ 
is chosen such that $|\Delta\theta_{i1}(t_{i1}+0)| 
<\pi/2$. (ii) At the second flip time  $t_{i2}$ we set   
\be 
|\Delta\theta_{i1}(t_{i2})|=2\pi/3.
\en  
For $t>t_{i2}$ we again introduce  
\be 
\Delta\theta_{i2} (t)=\Delta\theta_{i1}(t)  \pm \pi,
\en  
 where $+\pi$ or $-\pi$ 
is chosen such that $|\Delta\theta_{i2}(t_{i2}+0)| 
< \pi/2$. (iii)  We repeat   these 
procedures to obtain   the successive flip times.
 See  Fig.4 for    examples of the flip time series.
 
Note that the threshold $2\pi/3$ in Eqs.(A1) and (A3) 
may be changed to another angle, say $5\pi/6$. 
However,  the resultant flip time series is  
rather insensitive to 
this choice as long as it is in the range $[\pi/4,\pi/2]$.


\begin{thebibliography}{0}

\bibitem{Ang}
C. A. Angell, K. L. Ngai, G. B. McKenna, P. F. McMillan, and S. W. Martin, 
J. Appl. Phys. {\bf 88}, 3113 (2000).

\bibitem{Binder} 
K. Binder and W. Kob, {\it 
Glassy Materials and Disordered Solids} 
(World Scientific, Singapore, 2005).


\bibitem{Silescu} 
F. Fujara, B. Geil, H. Sillescu, and G. Fleischer, Z. Phys. B
{\bf 88}, 195 (1992); 
 M. T. Cicerone and M. D. Ediger, J. Chem. Phys. {\bf 104},
7210 (1996).


\bibitem{Rolf} 
A. Winkler, A. Latz, R. Schilling, and C. Theis, 
Phys. Rev. E {\bf 62},  8004 (2000).
 
\bibitem{Gotze} S.-H. Chong and W. G$\ddot{\rm{o}}$tze, 
Phys. Rev. E {\bf 65}, 041503 (2002).

\bibitem{Sch} R. Zhang and K. S. Schweizer, 
J. Chem. Phys. {\bf 133}, 104902 (2010); 
ibid. {\bf 136}, 154902 (2012).
\bibitem{ori} 
U. T. H\"{o}chli, 
K. Knorr, and A. Loidl, Adv. Phys. {\bf 39}, 405 (1990).


\bibitem{China} 
Z.  Zheng, F. Wang, and Y.  Han, Phys. Rev. Lett. 
{\bf 107}, 065702 (2011).

\bibitem{Mishra} 
C. K. Mishra, A. Rangarajan, and R. Ganapathy, 
Phys. Rev. Lett. {\bf 110}, 188301 (2013).

\bibitem{Klein} 
P. Sindzingre and M. L. Klein, J. Chem. Phys. {\bf 96}, 4681 (1992).

\bibitem{Lewis} L. J. Lewis and G. Wahnstr$\ddot{\rm o}$m, 
 Phys. Rev. E {\bf 50}, 3865 (1994);
J. Non-Cryst. Solids {\bf 172}, 69 (1994).

\bibitem{Debe} 
T. G. Lombardo, P. G. Debenedetti, and F. H. Stillinger,
J. Chem. Phys. {\bf 125}, 174507 (2006).


 
\bibitem{Kob1} S. K$\ddot{\rm a}$mmerer, W. Kob, and R.  Schilling, 
Phys. Rev. E {\bf 56},  5450 (1997). 

\bibitem{Michele} 
C. De Michele and D. Leporini, Phys. Rev. E {\bf 63}, 
 036702 (2001).
 

\bibitem{Chong1} 
S.-H. Chong, A. J. Moreno, F. Sciortino, and W. Kob,
Phys. Rev. Lett. {\bf 94}, 215701 (2005).
\bibitem{Moreno} 
A. J. Moreno, S.-H. Chong, W. Kob, and F. Sciortino, 
J. Chem. Phys. {\bf 123}, 204505 (2005).


\bibitem{Chong} 
S.-H. Chong and W. Kob,
Phys. Rev. Lett.  {\bf  102}, 025702 (2009). 





\bibitem{EPL} K.  Takae and A.  Onuki, 
EPL  {\bf 100},  16006 (2012).




\bibitem{Is} J. N. Israelachvili,  
{\it Intermolecular and Surface 
Forces} (Academic Press, London, 1991).


\bibitem{water} 
C. A. Angell and E. J. Sare, J. Chem. Phys. {\bf 49}, 
4713 (1968); 
 M. Kobayashi and H. Tanaka,   J. Phys. Chem. B,  {\bf 115}, 14077  (2011).

\bibitem{Lek} 
G. A. Vliegenthart, A. van Blaaderen, and H. N. W. 
Lekkerkerker, 
Faraday Discuss. {\bf 112}, 173 (1999). 
\bibitem{An} D. Antypov and D. J. Cleaver, 
J. Chem. Phys. {\bf 120}, 10307 (2004). 




\bibitem{Hamanaka} T. Hamanaka and A. Onuki, 
Phys. Rev. E 
{\bf 74}, 011506 (2006);  ibid. 
 {\bf 75}, 041503 (2007).

\bibitem{Shintani}
H. Shintani and 
H. Tanaka, Nat. Phys. {\bf 2}, 200 (2006).

\bibitem{Tanaka} 
T.  Kawasaki, T. Araki, and H. Tanaka,  Phys. Rev. Lett. 
{\bf  99}, 215701 (2007). 
\bibitem{YO} R. Yamamoto and A. Onuki,
J. Phys. Soc. Jpn. {\bf 66},  2545 (1997);  
 Phys. Rev. E {\bf 58}, 3515 (1998).   

\bibitem{Donati} 
C. Donati, J. F. Douglas, W. Kob, S. J. Plimpton, P. H. Poole, 
and S. C. Glotzer,  Phys. Rev. Lett. {\bf 80},2338 (1998).
\bibitem{Glotzer} 
S. C. Glotzer, J. Non-Cryst. Solids {\bf 274}, 342 (2000). 


\bibitem{Shiba}
H. Shiba, T. Kawasaki, and A. Onuki, 
 Phys. Rev. E {\bf 86}, 041504 (2012).



\bibitem{Gay}
J. G. Gay and B. J. Berne, J. Chem. Phys. {\bf 74}, 3316 (1981).


\bibitem{water2D} A. Ben-Naim,  J. Chem. Phys. {\bf  54}, 3682 (1971); 
K. A. T. Silverstein, 
 A. D. J. Haymet, and K. A. Dill, 
J. Am. Chem. Soc.  {\bf 120}, 3166 (1998). 
\bibitem{Leibler} 
J. M. Drouffe,  A. C. Maggs and S. Leibler, 
Science {\bf 254}, 1353 (1991);   
H. Noguchi, J. Chem. Phys. {\bf 134}, 055101 (2011).

\bibitem{nose}
S. Nos\'e, Mol. Phys. {\bf 52}, 255 (1984). 


\bibitem{Nelson}  D. R. Nelson  and B. I. Halperin, 
Phys. Rev. B  {\bf 19}, 2457 (1979). 


\bibitem{Pro} 
H. G. E. Hentschel, 
 V. Ilyin,  N. Makedonska, I. Procaccia,
 and N. Schupper,  Phys. Rev. E {\bf 75}, 050404(R) (2007).



\bibitem{Kawasaki-OnukiPRE}  T. Kawasaki and A. Onuki, 
 Phys. Rev. E {\bf 87}, 012312 (2013). 
This paper shows that   $D\tau_b$ 
is nearly independent of $T$ 
in a  supercooled  fragile binary mixture 
in three dimensions, 
where  $\tau_b$ is  the bond breakage 
time \cite{YO}. 



\bibitem{Ren} 
S. Sarkar,  X. Ren, and K. Otsuka, 
Phys. Rev. Lett. {\bf 95}, 205702 (2005); 
Y. Wang, X. Ren, and K. Otsuka,  
Phys. Rev. Lett. {\bf 97}, 225703 (2006).  


\bibitem{Cowley} 
R. A. Cowley,  S.N. Gvasaliya, 
S.G. Lushnikov, B. Roessli,  and G.M. Rotaru,  
 Adv. Phys.  {\bf  60}, 229 (2011). 


\end{thebibliography}
\end{document}